\documentclass[prb%
,twocolumn%
,floatfix%
,superscriptaddress%
,showpacs]{revtex4}
\usepackage{graphicx}
\usepackage{pstricks}
\usepackage{xspace}

\newcommand{\MATRIX}[4]{
\left(\begin{array}{cc}#1&#2\\#3&#4\end{array}\right)
}
\newcommand{\TILDE}[1]{\tilde{#1}}
\newcommand{\NN}{{\ensuremath{N}} }
\newcommand{\Nn}{{(N)}}
\newcommand{\Avar}{{\ensuremath{a}} }% barrier thickness
\newcommand{\LL}{{\ensuremath{L}} }  % barrier distance
\newcommand{\UU}{{\ensuremath{U}} }
\newcommand{\GG}{{\ensuremath{\frac{e^2}{h}}}}
\newcommand{\ef}{{\ensuremath{E_{\scriptscriptstyle F}}}}
\newcommand{\ek}{{\ensuremath{E_{\bf k}}}}
\newcommand{\kpar}{{\ensuremath{k_\parallel}}\xspace}
\newcommand{\Kpar}{{\ensuremath{{\bf k}_\parallel}}\xspace}
\newcommand{\DTK}{d^2\Kpar\;}
\newcommand{\Vol} [1]{{\bf #1}}
\newcommand{\Meq}[1]{eq.~(\ref{#1})}
\newcommand{\Mfactor}{.97}
\renewcommand{\Mfactor}{.7}
\renewcommand{\Mfactor}{1}
\newlength{\M}
\newlength{\T}
\begin{document}
%
%\setkeys{Gin}{draft=true}
\setkeys{Gin}{clip=true}
\title{
Landauer conductance of tunnel junctions:
Strong impact from boundary conditions
%Influence of boundary conditions
}
\author{Peter Zahn}%
\email{Zahn@physik.uni-halle.de}
\affiliation{Fysiska Institutionen, Uppsala Universitet, 
SE-75121 Uppsala, Sweden} 
\altaffiliation{
present address:
Fachbereich Physik, 
Martin-Luther-Universit{\"a}t Halle-Wittenberg, D-06099 Halle, Germany}%
\author{Ingrid Mertig}%
\affiliation{Fachbereich Physik, Martin-Luther-Universit{\"a}t Halle-Wittenberg,
D-06099 Halle, Germany}%
\date{\today}
\begin{abstract}
We present model calculations for the Landauer conductance of
tunnel junctions.
The tunneling of free electrons through a rectangular potential
barrier is considered. The conductance of a finite number of barriers was
calculated using a transfer matrix method. The periodic arrangement of
the same barriers
was described by a Kronig-Penney model to calculate the band structure
and from that the conductance of a point contact 
in the ballistic limit.
The comparison of both results elucidated the importance of the boundary
conditions. Caused by resonant
scattering in the superlattice the conductance is overestimated by the
order of 1/t, the transmission coefficient of the single barrier.
In the case of metallic multilayers these interferences are of minor
importance. 
In conclusion, the application of the Landauer formula to
periodic lattices to describe the tunneling conductance of a single
barrier is not correct.
%In the metallic regime the differences are quite small and the
%periodic lattice approach is applicable to point contacts of metallic
%systems.
\end{abstract}
%\Mmulti
\maketitle
\section{Introduction}
The discovery of Giant Magnetoresistance \cite{baibich88,binasch89} in
1988 initiated a renewed interest in the phenomenon of spin dependent
tunneling. First Julliere in 1975 \cite{julliere75} succeeded
to measure
the spin polarized current in a tunnel junction composed of two
ferromagnetic layers separated by an insulating barrier. Using improved
experimental techniques it is now possible to produce high quality
tunnel junctions showing a remarkable tunneling
magnetoresistance \cite{moodera95,miyazaki95,han01,leclair00}.
A widely used model to interpret the results in terms of the
polarization of the ferromagnetic leads was proposed by Julliere
\cite{julliere75}. It was based on earlier studies of
Meservey and Tedrow on tunneling between
superconductors and ferromagnets \cite{%tedrow71,tedrow73,
meservey94}.
The model assumed that  the tunnel current is proportional to the
product of the effective spin-dependent tunneling density 
for both ferromagnetic leads.
Using parabolic band models the influence of the barrier height and
width could be included in the theoretical investigations
\cite{slonczewski89,maclaren97}.

The electronic structure of the magnetic electrodes is, however, quite
different from the free-electron picture. This could be incorporated in the
considerations using tight-binding model hamiltonians 
\cite{%tsymbal99,
tsymbal01,mathon01}. Provided by the simplicity and
less demanding numerical effort of these models the influence of
defects on the tunneling current could be considered, too.
The up to now most sophisticated models use the self consistent
potential and electronic eigenstates of the whole tunnel junction to
determine the transport properties 
\cite{%maclaren98,
butler01,uiberacker01,wunnicke02}.
A well established formalism to treat tunneling conductance follows
the early ideas of Landauer and B\"uttiker 
\cite{
landauer57,%landauer70,
buettiker85}. The crucial property to
determine is the tunneling probability of the reservoir states through
the barrier. It is mostly calculated using the
Kubo-Greenwood approach based on Green's functions or a multiple
scattering formalism based on the electronic eigenstates.
In the experiments the thickness of the electrodes is much larger
than the electron mean free path representing homogenous
reservoirs and justifies the assumption made by Landauers formalism. 
Assuming semi-infinite boundary conditions for the electrode
hamiltonian the theoretical transport description is very close to the
experimental conditions.
Density functional theory based self-consistent electronic 
structure codes assume sometimes a three-dimensional periodicity of the system.
This means that the tunnel junction geometry is replaced by a periodic
system consisting of electrodes of finite thickness separated by
tunnel barriers with thicknesses used in the experiment. 
The implications for the self consistent potential and charge
densities are small due to the small screening length in the metallic
regions.
Considering metallic multilayers
Landauers formula was applied to the Co/Cu(001) system 
\cite{bauer94,schep95}.
In this case the theory describes the current through a (peculiar) point
contact made of a thin pillar and electrodes both with the same
periodic superstructure in current direction.

This paper is aimed at the importance of the boundary conditions for the
description of transport properties of tunnel junctions. 
The initial
idea was to determine starting from the eigenstates of a periodic array of
junctions the current through one of them in a 
geometry with infinite leads (electrodes) on both sides. 
The forthright attempt would be to adapt Landauers formula 
to the periodic lattice geometry assuming perfect transmission for all Bloch
states \cite{bauer94}. 
This attempt assumes a perfect contact of the barrier to reservoirs
with electronic states which are caused by the periodic lattice. The
conductance is determined by the contact resistance only \cite{sharvin65}.
One could suppose that the superlattice
quantization caused by the reflection at the barriers should
comprise the transmission probability occurring in the Landauer formula
for a single barrier.
By this assumption the determination of 
%As mentioned above the crux is the determination of 
the transmission
probability of the (semi-infinite) electrode eigenstates would be avoided.
The discussion later on will show the limitations of this approach and
the failure in the case of tunnel junctions. To determine the
conductance in the tunneling regime the determination of the
transmission probability is imperatively necessary. To extract these
quantities from a periodic lattice calculation a so-called inverse
Kronig-Penney-model was analyzed. 
The reversibility of them could be to demonstrated
obtaining the scattering properties of one barrier from the
eigenstates of the periodic system by means of a transfer matrix
formalism \cite{riedel01}. 

Using a parabolic band model a piece wise constant potential
in z-direction was assumed. The transition from a metallic to an insulating 
junction was studied by adjusting the Fermi level with
respect to the barrier height.
Similar free electron models have been used by several authors to
describe successfully the
transport properties of magnetic multilayers 
\cite{%bauer92,
hood92,visscher94,levy94,brataas94,vedyayev97}.
Starting from one rectangular barrier we will evaluate the transmission
probability and the resulting conductance of the free-electron-like 
reservoir states through a
finite number \NN of barriers. In the limit of large numbers \NN a
coincidence with the properties of the periodic system could be
expected, but could not be approved. 
We will discuss the implications of the periodic boundary
conditions on the transport properties and the limitations of the
periodic lattice approach.
The assumption about the shape of the potential are not important at
all for the general conclusions about the applicability of Landauer's
formula. 
\section{Model and transport equations}
\begin{figure}% fig1
\begin{center}
\begin{minipage}{\Mfactor\linewidth}
\M .05\linewidth
\T .05\M
\input{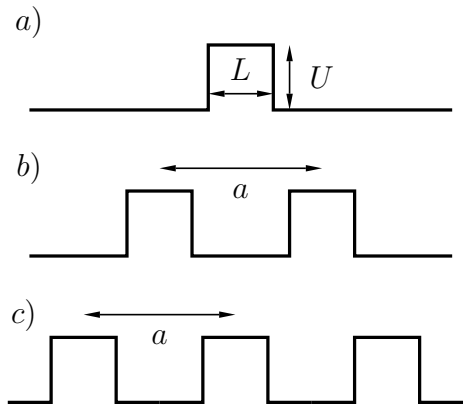}
\caption{
Geometrical arrangement of rectangular barriers of height $\UU$ and
width $\LL$, a) single barrier case, b) finite number $\NN$ of
barriers (shown for $\NN=2$) with a center-center distance of $\Avar$,
and c) periodic lattice
} 
\label{fig_geom}
\end{minipage}
\end{center}
\end{figure}  
To emphasize the influence of the boundary conditions we have chosen a
model system for the tunneling geometry. 
The 3-dimensional potential was chosen to be piece-wise constant.
The potential inside the barrier was set to $U$, outside it was set to
zero. The modulation of the potential is along the z-direction. The
thickness of the barrier was fixed to $L$. The single barrier case is
sketched in Fig.\ref{fig_geom}a).
A finite stack of barriers was
simulated by \NN barriers of width \LL with a constant distance \Avar,
and zero potential in the valley regions between the barriers and on
both sides as shown in Fig.\ref{fig_geom}b).
The barrier height $U=.38Ry$ and width $\LL=16.9 \text{a.u.}$ were
fixed, if not otherwise stated. 
In the periodic lattice the arrangement of barriers is translational
invariant in z-direction.
Emphasizing the periodicity of a system throughout this paper we
refer to the configuration in current direction which is the
z-axis. For the directions
perpendicular to the current a constant potential is assumed.
So, the momentum perpendicular to the z-axis \Kpar is a good quantum
number and an effective 
one-dimensional problem is to be solved for every \Kpar.

First, we consider a finite number \NN of barriers on both
sides in contact with free electron reservoirs with zero potential
and equal Fermi level \ef. 
The conductance in zero bias limit is obtained using Landauer's
formula \cite{landauer57} adding up all states at the Fermi level
\begin{eqnarray}
%\begin{equation}
G^\Nn &=& \GG \sum_{\kpar^2\le\ef} T^\Nn(\ef-\kpar^2)
\label{eq_gn_i}
, \\
T^\Nn(\ef-\kpar^2) &=& \left| \TILDE{t}^\Nn(\ef-\kpar^2) \right|^2
\quad .
%
%
%\end{equation}
\end{eqnarray}
{\GG } is the conductance quantum, $T^\Nn(\ef-\kpar^2)$ denotes the transmission
probability of state \Kpar with an energy equal $\ef$ derived from the
complex transmission coefficient $\TILDE{t}^\Nn$. 
Due to the simple potential shape the transmission coefficient depends
on the kinetic energy in z-direction only.
It was assumed that both reservoirs are
identical and a single channel exists for a given $\kpar$. 
The factor 2 for spin degeneracy will be omitted throughout this letter.
The neglection of the reflection probability with respect to the
primary formulation of Landauer corresponds to the
conductance value obtained by a two point measurement
\cite{buettiker85,sivan86}. The
difference to four point probe experiments arises from the contact
resistance \cite{sharvin65} and is negligible if the transmission T is
much smaller 
than unity. In the metallic regime for T near to unity this difference
can outweigh the calculated part of the conductance.
Our considerations are mainly focused on the tunneling regime.

Setting $A$
the cross section of the point contact the sum in
\Meq{eq_gn_i} can be transformed
\begin{equation}
G^\Nn = \GG \frac{A}{4\pi^2}
\int_{\kpar^2\le\ef} \DTK  T^\Nn(\ef-\kpar^2)
\label{eq_gn_ii}
\end{equation}
which can be written as an energy integral using the rotational symmetry of \Kpar 
$\DTK = 2\pi d\kpar\kpar$ and $E=\ef-\kpar^2$ 
\begin{equation}
G^\Nn = \GG \frac{A}{4\pi}
\int_0^\ef dE \;  T^\Nn(E)
\quad .
\label{eq_gn_iii}
\end{equation}
It is Evident, that the behavior of the conductance in dependence of
the filling of 
the reservoirs, given by \ef relative to the barrier height \UU, is simply
related to the transmission probability $T^\Nn(E)$ derived from a
one-dimensional model.
The transmission coefficient $\TILDE{t}^\Nn(E)$ will be calculated
using a transfer matrix formalism sketched in the next paragraph.

For a periodic arrangement of barriers we assume a perfect
transmission for all Bloch states. That is, we consider now a
periodic crystal described by a prolonged unit cell in z-direction
containing one barrier and one well region (super cell approach with
model potential). In the description of Landauer this corresponds to a
barrier in contact with reservoirs which have a modulated potential as
well. In this case the boundary conditions at both sides of the
barrier are not free electron like, and the states in the barrier
region have to match the eigenstates of the modulated reservoir
potential. 
Many ab initio electronic structure codes are adapted to a lattice
symmetry with 3-dimensional periodicity. So it seems to be desirable
to use these eigenstates to calculate the transport properties even in
the case of tunneling barriers. 
The conductance is given by an integral over the 3-dimensional Fermi
surface \cite{bauer94,schep95}
\begin{equation}
G^{SC} = \GG \frac{A}{4\pi^2} \frac{1}{2}
\int d^3{\bf k} \; \delta(\ek-\ef) \; v_{\bf k} {\bf n}
\quad  .
\label{eq_gsz_i}
\end{equation}
It is a sum enclosing the projections $S_\nu$ of the
Fermi surface sheets $\nu$ into the 
plane perpendicular to the transport direction ${\bf n}$
 (z-axis)
\begin{equation}
G^{SC} = \GG \frac{A}{4\pi^2} \frac{1}{2}
\sum_\nu S_\nu
\quad .
\label{eq_gsz_ii}
\end{equation}
To compute this integral for our model system the band structure has
to be evaluated at the Fermi level.
A Kronig-Penney-model from the textbooks was used and is sketched
briefly in the next paragraph.
\section{Scattering properties and Fermi surface}
The derivation of the transmission coefficients of a finite number of
barriers should be sketched briefly.
We start with a single barrier, and assume one incoming and one
outgoing solution on both sides of a 
symmetric barrier (Fig.~\ref{fig_geom}a). 
The wave function amplitudes on both sides entering the general solution of
the Schr\"odinger equation are connected by a 2x2 transfer matrix
\cite{merzbacher70}
\begin{equation}
M = \MATRIX{1/\TILDE{t}}{\TILDE{r}^*/\TILDE{t}^*}{\TILDE{r}/\TILDE{t}}
{1/\TILDE{t}^*}
\label{eq_m_matrix}
\end{equation}
with the complex transmission and reflection coefficients
\begin{eqnarray}
\TILDE{t}(E) &=& e^{-\imath k L} \left[ \cosh \kappa L + 
\frac{\imath}{2}\left(\frac{\kappa}{k} -\frac{k}{\kappa} \right)
\sinh \kappa L  \right]^{-1}
\nonumber\\
             &=& t(E) e^{\imath \delta(E)},
\label{eq_ct}
\\
\TILDE{r}(E) &=& - \imath \sqrt{1-t^2(E)} e^{\imath \delta(E)}
\label{eq_cr}
,
\end{eqnarray}
with $E=k^2$ and $U-E = \kappa^2$ and t(E) the absolute value of the
transmission coefficient.
%The transfer matrix $M$ couples the wave functions on one side of the
%barrier. 
The parameterization of M according to \Meq{eq_m_matrix} is
based on the assumption of a symmetric barrier ($V(z)=V(-z)$).

The total transmission coefficient of a stack of $\NN$
barriers $\TILDE{t}^\Nn (E)$ can be calculated by successive
application of the corresponding 
transfer matrices. The transfer matrix M of the single
barrier was derived for a symmetric barrier at $z=0$.
The transfer matrices of the other barriers in the stack have to be
transformed according to their shift in z-direction by 
appropriate D matrices. 
The transmission coefficient $t^\Nn$ depending on $E=k^2$ is obtained from
\begin{eqnarray}
M^\Nn &=& D^{(N-1)/2} \left(MD \right)^{(N-1)}M D^{(N-1)/2}\nonumber\\
&=&
\MATRIX{1/\TILDE{t}^\Nn}{\TILDE{r}^{\Nn*}/\TILDE{t}^{\Nn^*}}
{\TILDE{r}^\Nn/\TILDE{t}^\Nn}{1/\TILDE{t}^{\Nn*}},
\nonumber\\
\TILDE{t}^\Nn (E) &=& t^\Nn (E) e^{\imath \delta^\Nn(E)}, \text{and}
\nonumber\\
D &=& \MATRIX {e^{-\imath k a}}{0}{0}{e^{\imath k a}}
\quad .
\label{eq_m_n}
\end{eqnarray}
\begin{figure}
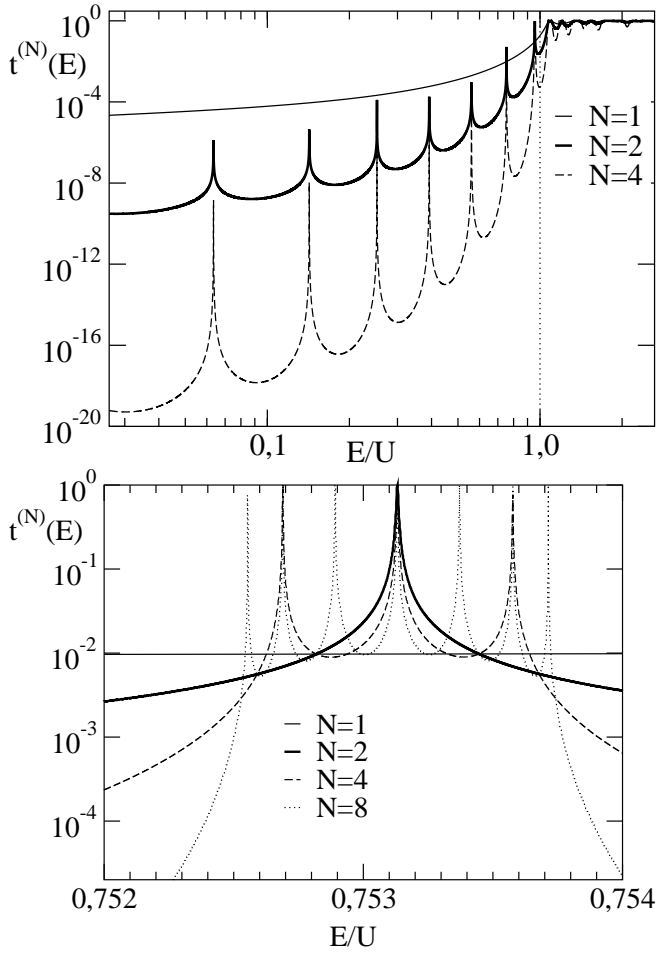
% fig2
\begin{center}
\begin{minipage}{\Mfactor\linewidth}
\includegraphics[width=\linewidth,clip=true]{fig2a.eps}
\includegraphics[width=\linewidth,clip=true]{fig2b.eps}
\caption{
Transmission coefficient $t^{(N)}(E)$ for stacks of $\NN$ rectangular
barriers
($U=.38\text{Ry}, \LL= 16.9 \text{a.u.}, \Avar = 54 \text{a.u.}$),
lower panel: behavior of $t^{(N)}(E)$ at resonance
} 
\label{fig_tn}
\end{minipage}
\end{center}
\end{figure}  
The behavior of $t^\Nn(E)$ is shown in
Fig.~\ref{fig_tn}. $t^1(E)=t(E)$ describes the single barrier case. For \NN
larger than one resonances occur for energies below $U$ corresponding to bound
states in the valley regions between the barriers. At the resonance the
transmission coefficient reaches unity in line with the
assumption of elastic scattering only. 
The maxima of the peaks are not covered by the chosen energy grid due
to their tiny width.
The implications
of additional inelastic scattering are discussed in the literature 
\cite{knauer77,ricco84,stone85}.
In the lower panel of Fig.~\ref{fig_tn} the behavior of the transmission next to a
resonance is shown on a finer energy mesh.
The number of sub-resonance peaks within one resonance interval increases
with the number of considered barriers.
To obtain $G^\Nn(\ef)$ an integration over $t^\Nn(E)$ from 0 to $\ef$
has to be performed according to eq. (\ref{eq_gn_ii}). 
The main contributions are given by the resonance
peaks which are a tiny fraction of the total interval (typical
$10^{-5}$ to $10^{-3}$). An adaptive mesh refinement algorithm similar
to Ref.~\cite{henk01} was developed to improve the Simpson integration
scheme up to the desired accuracy.

To obtain the Bloch states of a periodic arrangement of barriers,
sketched in Fig.~\ref{fig_geom}c), the Kronig-Penney model was used. The
energy bands are obtained from an implicit equation for the
perpendicular wave vector $\kappa$ of the superlattice 
\begin{equation}
\frac{\cos (k\Avar + \delta)}{t}  = \cos \kappa \Avar
\label{eq_kronig}
\end{equation}
which depends on the perpendicular wave vector \Kpar by 
$E=k^2=\ef-\kpar^2$, the transmission coefficient 
$t=t(E)$, and phase shift $\delta=\delta(E)$ of a single barrier
defined in \Meq{eq_ct}. 
The set of \kpar values providing a solution of
\Meq{eq_kronig} defines the projection of the Fermi
surface sheets required in \Meq{eq_gsz_ii}
\begin{equation}
S_\nu = 2 \pi (k^2_{\parallel \nu >} - k^2_{\parallel \nu <})
\quad .
\label{eq_fs_sheet}
\end{equation}
$k^2_{\parallel \nu <}$ and $k^2_{\parallel \nu >}$ are the inner and
outer radius of the annulus and are given by the condition $\frac{\cos
(k\Avar + \delta)}{t}  = \pm 1$. 
First the zeros and local maxima of $\cos(k\Avar+\delta)$ are
determined for $k \in (0,\sqrt{\ef})$, and the interval was partitioned in a
set of smaller ones. Every smaller interval providing one solution for
\Meq{eq_kronig} was treated by a scheme of nested intervals.
\section{Results and Discussion}
The behavior of $t^\Nn(E)$ is shown for several numbers \NN of
identical barriers in Fig.~\ref{fig_tn}. The upper panel shows the
over-all dependence 
including the basic solution $t(E)=t^{(1)}(E)$ for one barrier.
For energies larger than the barrier height \UU a nearly perfect
transmission is obtained with narrow anti-resonances with a reduced
transmission. 
For energies lower than the barrier height \UU the general trend of
$t^\Nn(E)$ is given by $[t(E)]^\NN$. Resonances with
perfect transmission occur at energies related to bound states
in the valleys between the barriers, see the lower panel of
Fig.~\ref{fig_tn}. 
For $\NN=2$ this corresponds to the 
results for double tunneling diodes without inelastic scattering
\cite{ricco83,ricco84,stone85,kalmeyer87}. Expanding $t^{(2)}(E)$
in the vicinity of the resonance energy $E_r$ a Lorentzian peak is
obtained 
\begin{eqnarray}
t^{(2)}(E) &=& t^2(E) \left| e^{2\imath(k \Avar + \delta(E))} + 
r^2(E)\right|^{-1}\nonumber \\
&=& \frac{\Gamma^2}{(E-E_r)^2 + \Gamma^2}
\label{eq_tn_res}
\end{eqnarray}
assuming a slowly varying $\delta(E)$ and $t=t(E) \ll 1$ in the
vicinity of $E_r$.
$E_r$ results from the condition $\cos(k(E)a+\delta(E))=0$.
The obtained value for $\Gamma = t^2k/(r\Avar)$ 
%with $t\approx t^{(1)}(E_r)$ 
is comparable to the 
resonance width induced by defect states in the middle of the barrier
in the case $\Avar-\LL\ll\Avar$ \cite{kalmeyer87}.
For $\NN>2$ barriers
$\NN-1$ peaks with perfect transmission appear around $E_r$. The
width of the total resonance is about 
\begin{equation}
\Delta_r = 4\sqrt{E_r}t/\Avar
\quad , 
\label{eq_width}
\end{equation}
which is of the order
$t^{-1}$ larger than $\Gamma$ and corresponds to the band width of the
Bloch states in the periodic system.
\begin{figure}% fig3
\begin{center}
\begin{minipage}{\Mfactor\linewidth}
\includegraphics[width=\linewidth]{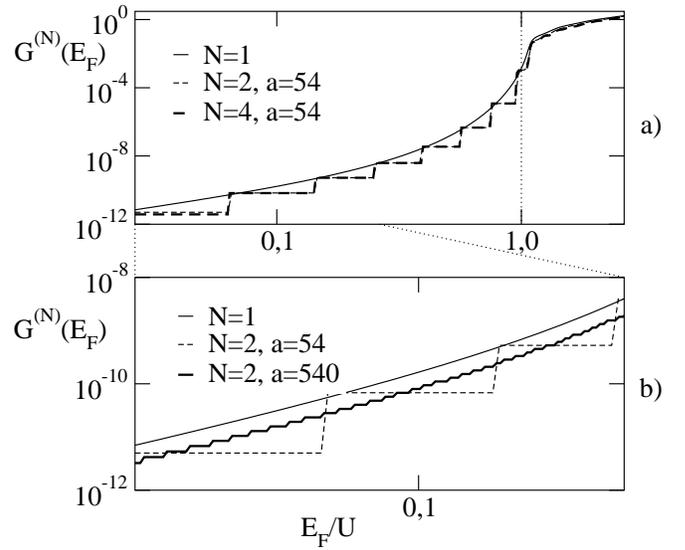}
\caption{
Landauer conductance $G^\Nn(\ef)$:
%($U=.38\text{Ry}$, $L=16.9 \text{a.u.}$, $\Avar=54 \text{a.u.}$), 
a) general trend, the curves for $\NN=2$ and $\NN=4$ nearly coincide,
b) $G^{(2)}(\ef)$ for different barrier distances $\Avar$ with
higher energy resolution.
} 
\label{fig_gn}
\end{minipage}
\end{center}
\end{figure}  

Fig.~\ref{fig_gn} shows the dependence of $G^\Nn$ as a function
of the Fermi level $\ef$. 
The numerical results are given in units of the prefactor $\GG
\frac{A}{4\pi^2}$ in eq.~\ref{eq_gn_ii}. 
The height of the barriers was fixed and small $\ef$
describes transport in the tunneling regime, but large $\ef$ with
respect to \UU models the (free electron like) metallic regime with
periodic potential perturbations. 
For small band fillings ($\ef\ll\UU$) the conductance increases
step-wise caused by the 
sharp resonance structure of the transmission coefficient $t^\Nn(E)$ for 
$\NN> 1$. 
With increasing Fermi level the resonance in $t^\Nn(E)$ appears in
k-space first at the
$\kpar=0$ point because the effective energy of the one-dimensional
tunneling problem $E=\ef-\kpar^2$ is the largest for these states. 
The spacing of the transmission resonances with respect to energy
scales with the inverse of the distance of the barriers $\Avar$ 
(see \Meq{eq_de}). 
As can be seen in
Fig.~\ref{fig_gn}b) for larger barrier distances the tunneling conductance of
$\NN>1$ follows exactly the trend of $G^{(1)}(\ef)$ apart from a factor of
the order 2.

The conductance contribution being added by one resonance in the
integral in \Meq{eq_gn_iii} can be
estimated in the case of small $t=t^{(1)}(E_r)$ by the product of the
averaged transmission probability $\overline{t^\Nn}^2$ and the width of the
resonance interval. From the analytical expression for
$t^\Nn(E)$ (for $t^{(2)}(E)$ see \Meq{eq_tn_res}) the mean
value of $\overline{t^\Nn}^2$ averaged over one resonance interval 
is nearly equal $\frac{\pi}{4}t$ independent on \NN.
% with tinier deviations for larger \NN . 
The total
width of the resonance interval is given by Eq. \ref{eq_width}.
%The spacing of the resonances is of the order
%$2\pi\frac{\sqrt{E}}{\Avar}$. 
Consequently the contribution of one
resonance is given by
\begin{equation}
\Delta G^{N>1} \approx
\frac{4\sqrt{E}t}{\Avar} \frac{\pi}{4}t = \frac{\pi\sqrt{E}}{\Avar}t^2
\quad .
\label{eq_gn_contrib}
\end{equation}
This is exactly half of the value obtained by the conductance 
integral for one barrier in \Meq{eq_gn_iii}
estimated by the resonance spacing 
\begin{equation}
\Delta E = 2 \Delta k\: k = 2 \frac{\pi}{\Avar} \sqrt{E}
\label{eq_de}
\end{equation}
times the transmission probability
${t}^2(E)$.
The resonance spacing is derived from the condition that the resonances
appear at energies of bound states in the valley region between
the barriers assuming hard walls ($\Delta k \: a = \pi$).
%The speculation was to reproduce $G^{(1)}(\ef)$ (disregarding the missing
%factor 2) if \NN is chosen sufficiently large.
This estimation is independent on the number \NN of barriers
and causes the same qualitative behavior of the conductance for a chosen
barrier distance $\Avar$, nevertheless the distance of the barriers
$\Avar$ determines the height and width of the resonance steps in
$G^\Nn(\ef)$, compare Fig.~\ref{fig_gn}b).
The cases $\NN=2$ and $\NN=4$ coincide exactly in Fig.~\ref{fig_gn}a.
\begin{figure}% fig4
\begin{center}
\begin{minipage}{\Mfactor\linewidth}
\includegraphics[width=\linewidth]{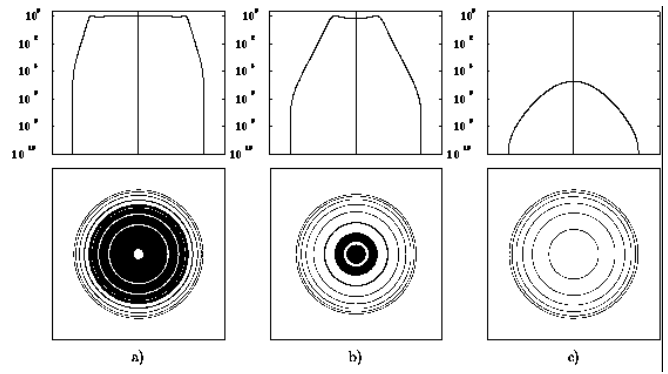}
\caption{
Transmission probability $T(\kpar)$ in
dependence on \kpar (upper row) and
supercell Fermi surface (lower row), $\ef=.25\text{Ry}$, 
barrier height $\UU$ is set $.38$ (a), $.76$ (b), and
$1.52$ (c) times $\ef$
} 
\label{fig_fs}
\end{minipage}
\end{center}
\end{figure}  

In the following attention is drawn to eigenstates, Fermi
surfaces and derived conductance integrals for the periodic lattice
containing the above described model tunneling barriers.
The Fermi surfaces of the periodic lattices are shown in
Fig.~\ref{fig_fs}. For reasons of 
comparability the Fermi level was fixed and the barrier
height was chosen to be $0.38$, $0.76$, and $1.52$ times $\ef$
representing the cases of a large, nearly equal and small Fermi energy
$\ef$ with respect to \UU, respectively. The upper row in
Fig.~\ref{fig_fs} shows the
dependence of the single barrier transmission 
$T^{(1)}=t^2(\ef-k_\parallel^2)$ on \kpar. 
%The shaded area marks the occupied states of the free electron
%reservoirs corresponding to the chosen Fermi energy.
The lower row shows the respective Fermi surfaces in a projection on
the plane perpendicular to the current direction. 
The radius of the outermost Fermi surface sheet corresponds to the
chosen Fermi level.
The area of the Fermi surface sheets projection is a
direct measure for the conductance contribution in \Meq{eq_gsz_ii}.
The strong relation
of transmission probability and the width of the Fermi surface sheets is
evident. For nearly perfect transmission free electron like sheets with
tiny gaps are obtained (left panel, central part). 
For small transmissions narrow sheets
occur (right panel). 
The width in direction of $\kpar$ can be estimated from
\Meq{eq_kronig} by linearizing
the cosine, assuming a slow variation of $t(E)$ and $\delta(E)$,
and the condition that the absolute value of the r.h.s. is smaller
than one 
\begin{equation}
\Delta k = \frac{2 t}{\Avar}
\quad.
\label{eq_kwidth}
\end{equation}
It corresponds to the energy width $\Delta_r$ of the
transmission resonance mentioned in \Meq{eq_width} above.
The center of the band is given by the condition 
$\cos (k\Avar + \delta)=0$ with $k^2=E=\ef-\kpar^2$.
\begin{figure}% fig5
\begin{center}
\begin{minipage}{\Mfactor\linewidth}
\includegraphics[width=\linewidth]{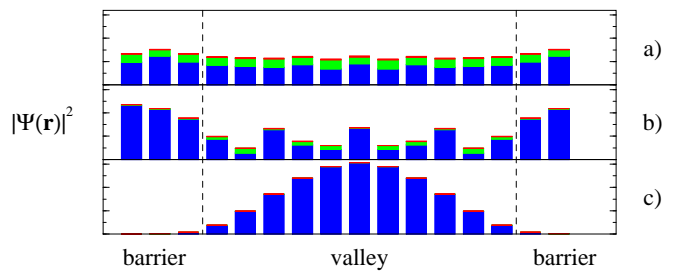}
\caption{
Probability amplitude profiles along the z-axis of typical eigenstates
of the periodic lattice, 
a) $\ef-\kpar^2>U$, 
b) $\ef-\kpar^2\approx U$, and 
c) $\ef-\kpar^2\ll U$
} 
\label{fig_ampl}
\end{minipage}
\end{center}
\end{figure}  

The scattering properties of the single barrier 
%the periodic system is assembled from 
are reflected by the eigenstates of the periodic system and can be
determined using an 'inverse' Kronig-Penney-model \cite{riedel01}.
Typical eigenstates at the Fermi level for different $\kpar$ are
characterized in 
Fig.~\ref{fig_ampl} by 
their probability amplitude profiles along the z-axis. 
$\ef$ was chosen to be larger then $U$ (Fig.~\ref{fig_fs}, left panel).
Fig.~\ref{fig_ampl}a) shows an eigenstate close to the center of
the Brillouin zone ($\kpar\approx 0$, $\ef-\kpar^2>U$). The probability
amplitude is slightly modulated by the periodic potential, because
the transmission of the barrier is nearly perfect ($t\approx 1$). For
states with a larger $\kpar$ the energy $\ef-\kpar^2$
decreases and Fig.~\ref{fig_ampl}b) shows a state close to
$\ef-\kpar^2=U$. A resonance inside the barrier is a typical feature
for these states with a moderate transmission coefficient ($t\alt
1$). States with a small effective energy $E=\ef-\kpar^2\ll U$ which
corresponds to large $\kpar$ are confined to the valley region between
the barriers, caused by the strong reflection ($t\ll 1$, see
Fig.~\ref{fig_ampl}c). 
\begin{figure}% fig6
\begin{center}
\begin{minipage}{\Mfactor\linewidth}
\includegraphics[width=\linewidth,clip=true]{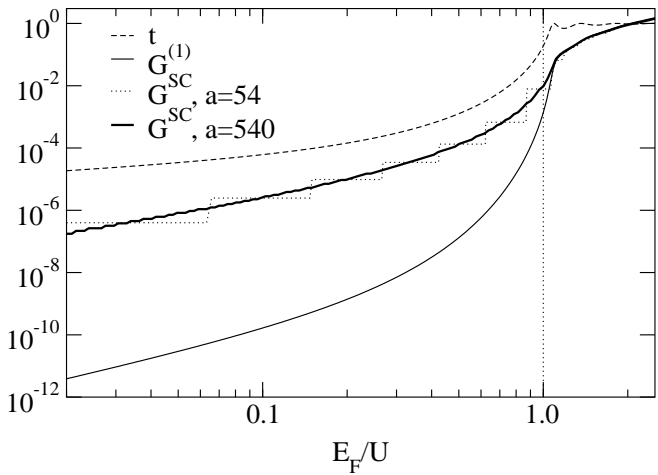}
\caption{
Conductance $G^{SC}$ in dependence on band filling of the
reservoirs ($\ef$), 
for comparison the exact result for a single barrier $G^{(1)}$ is given,
($U=.38\text{Ry}, \LL=16.9 \text{a.u.}$)
} 
\label{fig_gsz}
\end{minipage}
\end{center}
\end{figure}  

Figure \ref{fig_gsz} shows the main result of these investigations,
the dependence of the conductance $G^{SC}$ on the Fermi level applying
\Meq{eq_gsz_ii}. For comparison the exact result for the one barrier
conductance $G^{(1)}$ and the transmission coefficient $t(E)$ are
given. 
$G^{SC}$ is in good agreement with $G^{(1)}$ in the metallic regime
($\ef/U>1$). The restrictions caused by the
neglection of the 
reflection coefficients R in the Landauer formula discussed
above would rescale both curves in the same manner \cite{sivan86}.
In the tunneling regime ($\ef \ll U$) the supercell conductance $G^{SC}$
overestimates the correct value by the order of 
inverse t, as it is visible in the logarithmic plot. 

The discrepancies can be understood comparing  the corresponding
contributions to $G^{(1)}$ in \Meq{eq_gn_iii} and to $G^{SC}$ in
\Meq{eq_gsz_ii} for one resonance interval and the corresponding Fermi
surface sheet, respectively. 
One has to assume large $\Avar$, so that $t(E)$ and $\delta(E)$ do not change
significantly in a resonance interval.
A similar estimation as given above for $\Delta G^\Nn$ in
\Meq{eq_gn_contrib} can be done for 
$G^{SC}$ evaluating the contribution for a single Fermi surface sheet.
Disregarding the prefactors of the integral in
\Meq{eq_gsz_ii} the contribution of one Fermi surface sheet is given
by the circumference $2 \pi k=2 \pi \sqrt{E}$ times the width 
$\Delta k=\frac{2 t}{\Avar}$ from \Meq{eq_kwidth}
\begin{equation}
S_\nu = 2 \pi \sqrt{E} \; \frac{2t}{\Avar} \quad .
\label{eq_gsz_contrib}
\end{equation}
The corresponding contribution $\Delta G^{(1)}$ in \Meq{eq_gn_iii} is
given using \Meq{eq_de} by
\begin{equation}
\Delta G^{(1)}\Delta G^{(1)} =\Delta E \, t^2 =2\pi/a\sqrt{E} t^2 .
\end{equation}
Comparing these two expressions
including the prefactors in \Meq{eq_gn_iii} and
\Meq{eq_gsz_ii} the ratio of $G^{SC}$ and $G^{(1)}$ can be
estimated by $\frac{2}{\pi t}$. 
The main contribution to the integrals is given by the Fermi surface
sheet and transmission resonance interval, respectively, with the
largest effective energy $\ef-\kpar^2$ for small $\kpar$. So the ratio
$G^{SC}/G^{(1)}$ can be estimated by the transmission coefficient
$t(\ef)$ at the Fermi level.
A closer look to the
curves in Fig.~\ref{fig_gsz} reveals the prefactor $2/\pi t$.

Another picture to illustrate the differences of both approaches is the
following. 
Using the supercell approach a perfect transmission for all periodic
lattice Bloch states is assumed. In terms of the transmission
coefficient $t(E)$ in dependence on energy this means to
set $t^\Nn$ equal unity inside the resonance intervals for large
$\NN$.
(see Fig.~\ref{fig_tn}b), and to zero outside the resonance
regions. 
The resonance region for large $N$ give exactly the position of the
periodic lattice bands.
The differences outside the resonances are negligible but
the contribution of the resonances will be overestimated by far, as
given by $\overline{t^\Nn}$. One
has to keep in mind the logarithmic scale in Fig.~\ref{fig_tn}b) and
the small width of the $\NN-1$ sub-resonance peaks.
\section{Conclusions}
Based on calculations of the Landauer conductance for tunneling
barriers in different geometrical configurations the influence of the
boundary conditions could be elucidated. 
Comparing the conductance of one and a finite number of identical
rectangular barriers we found the same qualitative behavior in the dependence on
the band filling of the electrodes (Fermi level) with respect to the
barrier height $U$. For Fermi
levels smaller than the barrier height (tunneling regime) both results
coincide despite a factor of 2 even for large numbers $\NN$.
Using a supercell approach and calculating the conductance
from the periodic lattice eigenstates assuming perfect transmission, one overestimates the exact result
in the tunneling regime by order $t^{-1}$, the transmission
coefficient of the single barrier for states at $\kpar=0$.
In the tunneling regime the small transmission and the resulting
interferences of the states in the reservoirs prevent a description of the single
barrier conductance by a periodic lattice. The inclusion of the
correct semi-infinite boundary conditions for the 
eigenstates of the reservoirs is crucial.

In the metallic regime  ($\ef>U$) both results coincide and
the approach was already successfully applied to metallic multilayer point
contacts in CPP geometry \cite{schep95}.

We would like to thank J. Mathon for stimulating critics.
P.Z. likes to acknowledge financial support by the Swedish Foundation
for Strategic Research and fruitful discussions with Susanne Mirbt.
\end{document}